\documentclass[prb,aps,twocolumn,showpacs]{revtex4-1}

\usepackage{graphicx}
\usepackage{here}
\usepackage{bm,color}
\usepackage{physics}
\usepackage{amsmath}
\usepackage{bm,color}
\usepackage{multirow}

\newcommand{\be}{\begin{eqnarray}}
\newcommand{\ee}{\end{eqnarray}}

\newcommand{\Se}{S_{\rm ent}}
\usepackage{ulem}

\usepackage{xcolor}

\newcommand{\IP}{I_{\rm IPR}}

\def\W{h}

\begin{document}

\title{
Interplay and competition between disorder and flat band in an interacting Creutz ladder
}
\date{\today}
\author{Takahiro Orito}
\affiliation{Quantum Matter program, Graduate School of Advanced Science and Engineering, 
Hiroshima University, Higashi-Hiroshima 739-8530, Japan}
\author{Yoshihito Kuno}
\affiliation{Department of Physics, Graduate School of Science, Tsukuba University, Tsukuba, Ibaraki 305-8571, Japan}
\author{Ikuo Ichinose}
\affiliation{Department of Applied Physics, Nagoya Institute of Technology, Nagoya, 466-8555, Japan}

\begin{abstract}
We clarify the interplay and competition between disorder and flat band in  
the Creutz ladder with inter-particle interactions focusing on the system's dynamics. 
Without disorder, the Creutz ladder exhibits flat-band many-body localization 
(FMBL). 
In this work, we find that disorder generates drastic effects on the system, i.e.,
addition of it induces a thermal phase first and further increase of it 
leads the system to the conventional many-body-localized (MBL) phase. 
The competition gives novel localization properties and 
unconventional quench dynamics to the system. 
We first draw the global sketch of the localization phase diagram by focusing 
on the two-particle system. 
The thermal phase intrudes between the FMBL and MBL phases, the regime of which 
depends on the strength of disorder and interactions. 
Based on the two-particle phase diagram and the properties of the quench dynamics, 
we further investigate finite-filling cases in detail. 
At finite-filling fractions, we again find that the interplay/competition between
the interactions and disorder in the original
flat-band Creutz ladder induces thermal phase, 
which separate the FMBL and MBL phases. 
We also verify that the time evolution of the system coincides with the 
static phase diagrams. 
For suitable fillings, the conservation of the initial-state information and 
low-growth entanglement dynamics are also observed. 
These properties depend on the strength of disorder and interactions.
\end{abstract}

\maketitle
\section{Introduction}
Localization of the quantum system
has been an important hot topic in the condensed matter 
physics for more than 50 years \cite{AL50}.
There, disorder plays a key role to make quantum states localized. 
It has been recognised that the properties of localization 
continue to exist even in the presence of inter-particle interactions \cite{Basko2006}.
This interacting localization phenomenon is called many-body 
localization (MBL) \cite{Nandkishore}.
The MBL phase is not an ergodic state, i.e.,
localized energy eigenstates break eigenstate thermalization 
hypothesis (ETH) \cite{Deutsch,Srednicki}.  
Related to the ETH breaking, the quench dynamics in MBL phases exhibits 
unconventional behavior, 
i.e., an initial non-entangled state does not get thermalized 
and logarithmic growth of entanglement entropy (EE) appears \cite{Abanin,Imbrie,Serbyn,Bardarson}.
The origin of MBL is the presence of the emergent local integrals of motion
(LIOMs) \cite{Huse,Bera,Imbrie1}, which give the system a kind of integrability. 
The LIOMs in the MBL system are composed of disorder-induced
dressed ``spin operators" called $\ell$-bit \cite{Huse}.  
The states created by different $\ell$-bits are orthogonal with each other,
and an effective Hamiltonian describing MBL is constructed solely in terms of
the number operators of the $\ell$-bit, the LIOMs.

Conventional MBL occurs in a disordered system. 
However, the MBL phenomenon is not limited to such a disordered system.
Indeed, the MBL-like phenomena in disorder-free interacting systems have been 
reported \cite{Smith1,Smith2,Surace,Paul,Li2021}. 
In such a translation-invariant system, the presence of the LIOMs is a key factor 
to understand MBL phenomena. 
In particular, among such disorder-free systems, completely flat-band systems 
such as the Creutz ladder \cite{Creutz1999,Bermudez,Junemann,Shin2020,kuno2020} and diamond lattice \cite{vidal0,vidal1,Mukherjee2018,Pelegri} 
have attracted a lot of attention.
In these systems without interactions, compact localized states (CLS) \cite{Leykam},
which are single-particle energy eigenstates in the flat bands, 
works as $\ell$-bits, and they induce localization called 
Aharanov-Bohm (AB) caging \cite{vidal0,vidal1}. 
Furthermore for complete flat-band systems, even after including inter-particle
interactions, a CLS picture can survive and induce a disorder-free MBL, 
which one calls the flat-band MBL (FMBL) \cite{KOI2020,Danieli_1}. 
Recently, higher-dimensional models exhibiting FMBL were constructed systematically~\cite{IchiOriKu}.
The detailed study on the nature of such FMBL states is an important topic now.
In particular concerning the $\ell$-bits in the above mentioned systems in random potentials, 
there are important and interesting problems:
whether dressed CLS emerge there and if so, 
whether the resultant LIOMs induce the FMBL properties.    
Some previous studies on the above problems have been reported to
confirm the existence of FMBL-like states in various flat-band based models \cite{Roy,Zurita,Daumann,Tilleke,Khare,Liberto,Danieli1,Danieli2,OKI2020}.

So far we have investigated the physical properties of FMBL 
in the Creutz ladder by focusing on 
some limited parameter regimes \cite{KOI2020} and also the dynamical aspect 
without disorders \cite{OKI2021}.
In this work, beyond our previous studies \cite{KOI2020,OKI2021}, 
we shall systematically investigate 
the interplay and competition between the flat band and disorder. 
Such interplay and competition may induce further interesting and significant
MBL physics including counter-intuitive phenomena. 
In certain cases, adding disorders to a flat-band system generates
a {\it delocalized phase} \cite{Goda,Shukla2018}. 
It is interesting to investigate if such an exotic behavior occurs also
in a complete flat-band model with interactions. 
In this work, we first try to draw a localization phase diagram by 
considering a simple two-particle case. 
There, we find interesting global phase structure, i.e., a thermal phase 
intrudes between FMBL and MBL phases. 
A similar phase structure has been recently observed for certain disordered systems
with interactions; the existence of a thermal phase between two distinct MBL states
generated by quench disorders. 
In contrast to these works, our study concerns transitions between
FMBL and MBL, where the behavior of the LIOMs, constructed explicitly
in the original Creutz ladder, plays an important role.

Next, based on the two-particle results, we shall investigate systems with finite
particle densities.
For a system at a small filling fraction, we study the static properties concerning
localization and find that the system's phase structure is 
close to that of the two-particle system. 
Furthermore, we investigate the dynamical aspect, in particular, 
entanglement dynamics. 
We observe slow-growth of the entanglement entropy (EE) for a certain low filling.
Global properties of the system depends on the filling fractions, and 
ordinary phase diagram is observed for a system at relatively high fillings.

This paper is organized as follows.
In Sec.~II, we introduce the Creutz ladder in random potentials and also 
with interactions.
We briefly review its flat-band limit, in particular, the existence of 
the $\ell$-bits and phase diagram without disorder and interactions. 
In Sec. III, we study the two-particle systems of the Creutz ladder.
We calculate the the EE, the number EE, the expectation 
value of the fourth-moment of the LIOMs, and the time evolution of various 
quantities to obtain the phase diagram of the two-particle system.
As a result of interplay and competition between the flat-band nature and disorder
effects, there emerge three phases:
the FMBL, ETH and MBL in the (disorder)-(interaction) plane.
Section IV is devoted for the study on the Creutz ladder at finite fillings.
We first investigate static localization properties by calculating EE, 
the variance of EE, and the inverse participation ratio (IPR).
Phase diagram depends on the strength of the interactions, as in 
the two-particle system studied in Sec.~III.
By using the schematic phase diagram obtained by the static properties of the system 
as a guide, we focus on some specific parameter ranges, and study
the time evolution of the EE and the return probability. 
Studies in Sec.~IV reveal that the phase diagram of the system substantially depends
on the filling fraction; i.e., at $1/6$ filling, there emerge three phases 
as in the two-particle system,
whereas at $1/4$ filling, the system has more `standard' properties concerning
the ETH and MBL.
Section V is devoted for conclusion and discussion.

\begin{figure}[t]
\begin{center} 
\includegraphics[width=6cm]{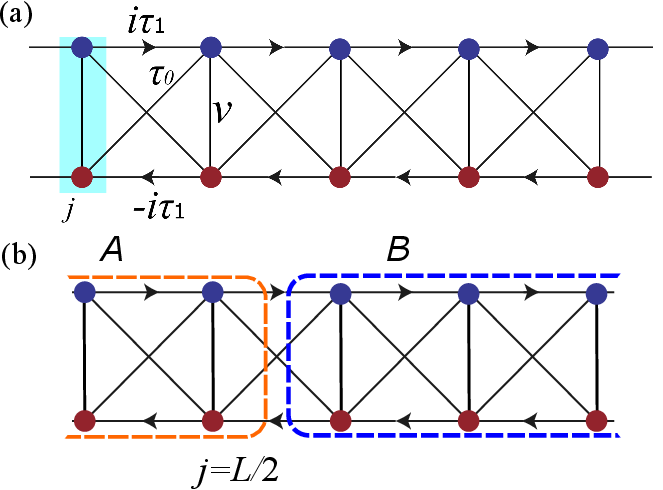}  
\end{center} 
\caption{(a) Schematic figure of the Creutz ladder. The blue shade regime 
represents the unit cell.
The blue and red sites are $a$ and $b$ sub-lattice where the operators 
$a_j$ and $b_j$ reside. 
(b) Entanglement cut for the calculation of the EE.
}
\label{model}
\end{figure}
\section{Interacting Creutz ladder in random potentials}

\begin{figure*}[t]
\begin{center} 
\includegraphics[width=12cm]{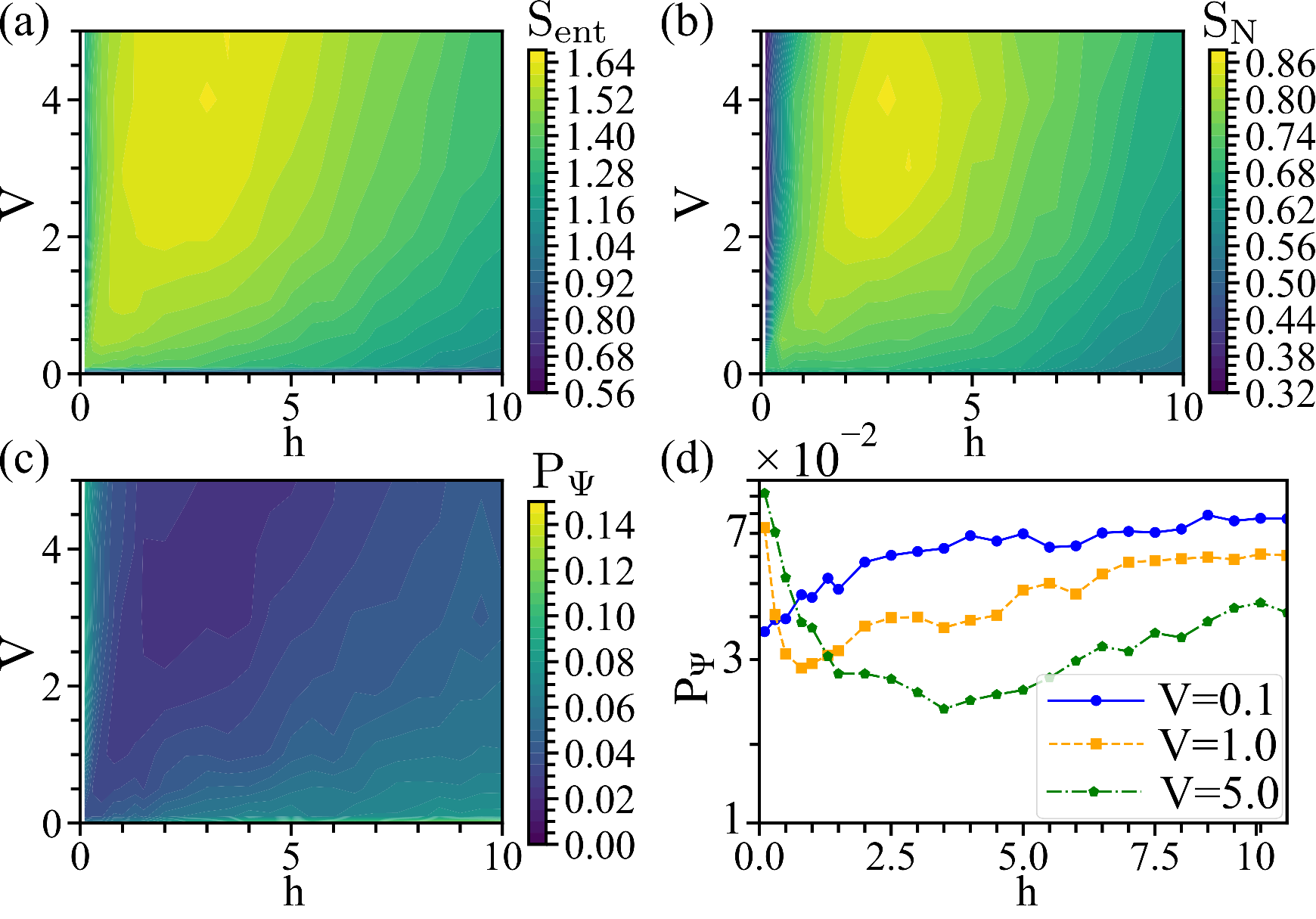}  
\end{center} 
\caption{$(\W-V)$-plane phase diagram obtained by some physical quantities in time evolution of two-particle system.
We plot the values at the final stage,$t> 10^{3}$.
(a) EE, $\Se$ (b) Number EE, $S_N$ (c) FMOL, $P_\psi$.
All data indicate the existence of two kinds of localized phases 
(FMBL and MBL) and thermal phase between them.
(d) For $V=0.1$, $P_\psi$ is a monotonous increasing function of $\W$, 
which indicates 
the FMBL state disappears by the weak interaction.
On the other hand for $V=1.0$ and $5.0$, $P_\psi$ decreases first in the small $\W$ regime, and then 
it starts to increase as $\W$ increases.
The FMBL state exists for the small $\W$ regime, the thermal state takes the place 
of it for 
intermediate $\W$, and then the state of ordinary MBL appears for large $\W$.
For data, the linear system size $L=10$.
}
\label{Fig1s}
\end{figure*}

Target model is the Creutz ladder whose Hamiltonian is given as follows 
(See also Fig.~\ref{model} (a) for its schematic figure),
\be
H_{\rm CL} = && \sum^{L}_{j=1}\Big[-i\tau_1 (a^\dagger_{j+1}a_j-b^\dagger_{j+1}b_j) \nonumber \\
&&\hspace{1cm} -\tau_0(a^\dagger_{j+1}b_j+b^\dagger_{j+1}a_j)+ \mbox{h.c.} \Big],
\label{HCL}
\ee
where $a_j$ and $b_j$ are fermion annihilation operators at site $j$, and $\tau_1$ and $\tau_0$ are
intra-chain and inter-chain hopping amplitudes, respectively. 
The corresponding system has been experimentally realized \cite{Shin2020}.
For $\tau_1=\tau_0$, the system has two complete-flat bands, 
and perfect localization called the AB caging takes place there.
Hereafter, we set $\tau_{0}=\tau_{1}$. 
Then, the Hamiltonian (\ref{HCL}) is expressed in terms of LIOMs
with $\ell$-bits, $\{W^\pm_j\}$, residing on plaquettes of the ladder and defined by,
\be
W^+_j&=&{1 \over 2}(-ia_{j+1}+b_{j+1}+a_j-ib_j),  \nonumber  \\
W^-_j&=&{1 \over 2}(-ia_{j+1}+b_{j+1}-a_j+ib_j).
\label{Wab}
\ee
Here, the flat-band Hamiltonian is given as: 
\be
H_{\rm flat} &\equiv& H_{\rm CL}|_{\tau_1=\tau_0}  \nonumber \\
&=& \sum^{L}_{j=1} [-2\tau_0W^{+\dagger}_j W^+_j+2\tau_0W^{-\dagger}_jW^-_j].
\label{HWW}  
\ee
From Eq.~(\ref{HWW}), the LIOMs are explicitly given by $K^\pm_j=W^{\pm\dagger}_jW^\pm_j$.
It is readily verified
$
\{W^{\alpha \dagger}_j ,W^{\beta}_k \}=\delta_{\alpha, \beta}\delta_{jk}, 
\;\; (\alpha, \beta = \pm), \; \mbox{etc.}
$
Then, it is obvious that all energy eigenstates of $H_{\rm flat}$ are obtained by 
applying the creation operators $\{W^{\pm\dagger}_j\}$ to
the vacuum state, and the single-particle spectra are given by $E=\pm 2\tau_0$.
Eigenstates are strictly localized in a single plaquette and the Hamiltonian (\ref{HWW}) describes a disorder-free 
and full-localized system.

By the specific choice of the coefficients of the hoppings in $H_{\rm CL}$ in Eq.~(\ref{HCL}), 
Eq.~(\ref{HWW}) is realized.
The Hamiltonian $H_{\rm CL}$ [Eq.~(\ref{HCL})] describes a system with non-trivial
magnetic flux.
By using local gauge transformations without changing the net magnetic flux 
and its pattern, 
the coefficients of the intra-chain hoppings can be converted
to real numbers while the inter-chain hopping comes to be imaginary. 

In this work, we study effects of interactions between the fermions as well as random local potentials,
both of which are given as follows:
\be
&& H_{\rm I}=V\sum_j (n^a_jn^b_j+n^a_jn^a_{j+1}+n^b_jn^b_{j+1}), \label{HI} \\
&& H_{\rm RP}= \sum_j (\mu^a_jn^a_j +\mu^b_jn^b_j),
\label{HRP}
\ee
where the number operators $n^{a(b)}_j=a^\dagger_ja_j (b^\dagger_jb_j)$, $V(>0)$ 
denotes the strength of the repulsive interactions,
and local random potentials have a uniform distribution such as, $\mu^{a(b)}_j \in [-\W,+\W]$ with a positive constant $\W$.
In the presence of the random potential $H_{\rm RP}$ in Eq.~(\ref{HRP}), 
energy eigenstates, which are strictly localized in a single plaquette in $H_{\rm flat}$, 
get to extend over few plaquettes.

In the rest of the present work, we consider the system $H_{\cal T}=H_{\rm flat}+H_{\rm I}+H_{\rm RP}$
and are interested in its phase diagram in the $(\W-V)$-plane.
For the case of $H_{\rm RP}=0$, the system $H_{\rm flat}+H_{\rm I}$ exhibits non-trivial
behaviors in the dynamics depending on the value of $V$.
As the parameter $V$ is getting large, single-particle states in multi-particle systems are deformed by the repulsion,
and some kind of oscillation in the LIOMs takes place.
However, the effects of the LIOMs survive there and the development of the EE is substantially suppressed.
Furthermore in Ref.\cite{KOI2020}, we also observed certain interesting interplay of $H_{\rm I}$ and $H_{\rm RP}$ with some fixed interactions.
That is, by studying the energy-level statistics and participation ratio, we found that
a transition from FMBL to ordinary MBL takes place as the value of $h$ is increased.
In the present paper, we shall investigate the interplay between $H_{\rm I}$ [Eq.~(\ref{HI})] and 
$H_{\rm RP}$ [Eq.~(\ref{HRP})] by observing dynamical properties of the system 
$H_{\cal T}$ in detail.


\section{Two-particle system}

First of all, we study time evolution of two-particle systems of $H_{\cal T}$, 
and draw their ``phase diagram''.
This study is motivated by the works of Ref. \cite{Serbyn2012_2}.
To this end, we measure time evolution of 
the half-ladder EE [denoted as $S_{\rm ent}$], and the fourth-moment of the LIOMs (FMOL).
The EE is defined as $S_{\rm ent}=-\mbox{Tr} [\rho_A \ln (\rho_A)]$ with $A$-subsystem reduced density matrix
$\rho_A=\mbox{Tr}_B[\rho]$. 
We furthermore calculate the number EE \cite{Lukin2019,Kiefer-Emmanouilidis} $S_N=-\sum_n p(n) \ln p(n)$, where $p(n)$ is the probability
of finding $n$ particles in the $A$-subsystem, and $S_{\rm ent}=S_N+S_c$, where $S_c$ is called configurational entropy.
On the other hand, the FMOL is defined as $P_\psi(t)=\sum_\pm P^\pm_\psi(t)$, 
where 
$P^{\pm}_\psi(t) \equiv \sum^{L}_{j=1}|\langle \psi(t)|K^{\pm}_j|\psi(t)\rangle|^4$
for a time-evolving state $|\psi(t)\rangle$.

As an initial state $|\psi\rangle_{W}$, we first consider
a two-particle system in which each particle state consists 
of a superposition of two states such as 
\be
|\psi\rangle_{W}
&&= 
{1 \over 2}(W^{+\dagger}_j-W^{-\dagger}_j)(W^{+\dagger}_k-W^{-\dagger}_k)|0\rangle  \nonumber \\
&&= {1\over 2}(a^\dagger_j+ib^\dagger_j) (a^\dagger_k+ib^\dagger_k)|0\rangle.
\label{a+ib}
\ee
For sufficiently small $V$ and $\W$, the LIOMs, $K^\pm_j$'s are stable and therefore
$P_\psi(t)$ is almost constant, whereas it exhibits some kind of unstable behavior as $V$ and/or $\W$ are
getting large because of deformation of the original LIOMs, $K^{\pm}_j$, in $H_{\rm flat}$.
In the practical calculation, we put $j=4$ and $k=6$ in Eq.~(\ref{a+ib}), and $A$-subsystem is the region 
of the ladder for sites $i<6$. [See Fig.\ref{model} (b). 
We take $(L=10)\times 2$ system size.] The probability of finding $n$ particle is calculated by using the obtained wave functions
of the full system. 
It is given by $|\Psi(t)\rangle=\sum_{n_A,k,\ell}\psi^{n_A}_{k,\ell}|(n_A,k)\rangle_A \otimes|(N-n_A,\ell)\rangle_B$, where $N$ and $n_A$ are the total particle number 
and the particle number of A-subsystem, $|(n_A,k)\rangle_A \otimes|(N-n_A,\ell)\rangle_B$
is a Fock state base with $n_{A}$ particles in $A$-subsystem and, $\psi^{n_A}_{k,\ell}$ 
is the coefficient.  
Then, $p(n)=\sum_{k,\ell}|\psi^{n}_{k,\ell}|^2$.
Ensemble average concerning the disorder realization is also taken.

\begin{figure}[t]
\begin{center} 
\includegraphics[width=7cm]{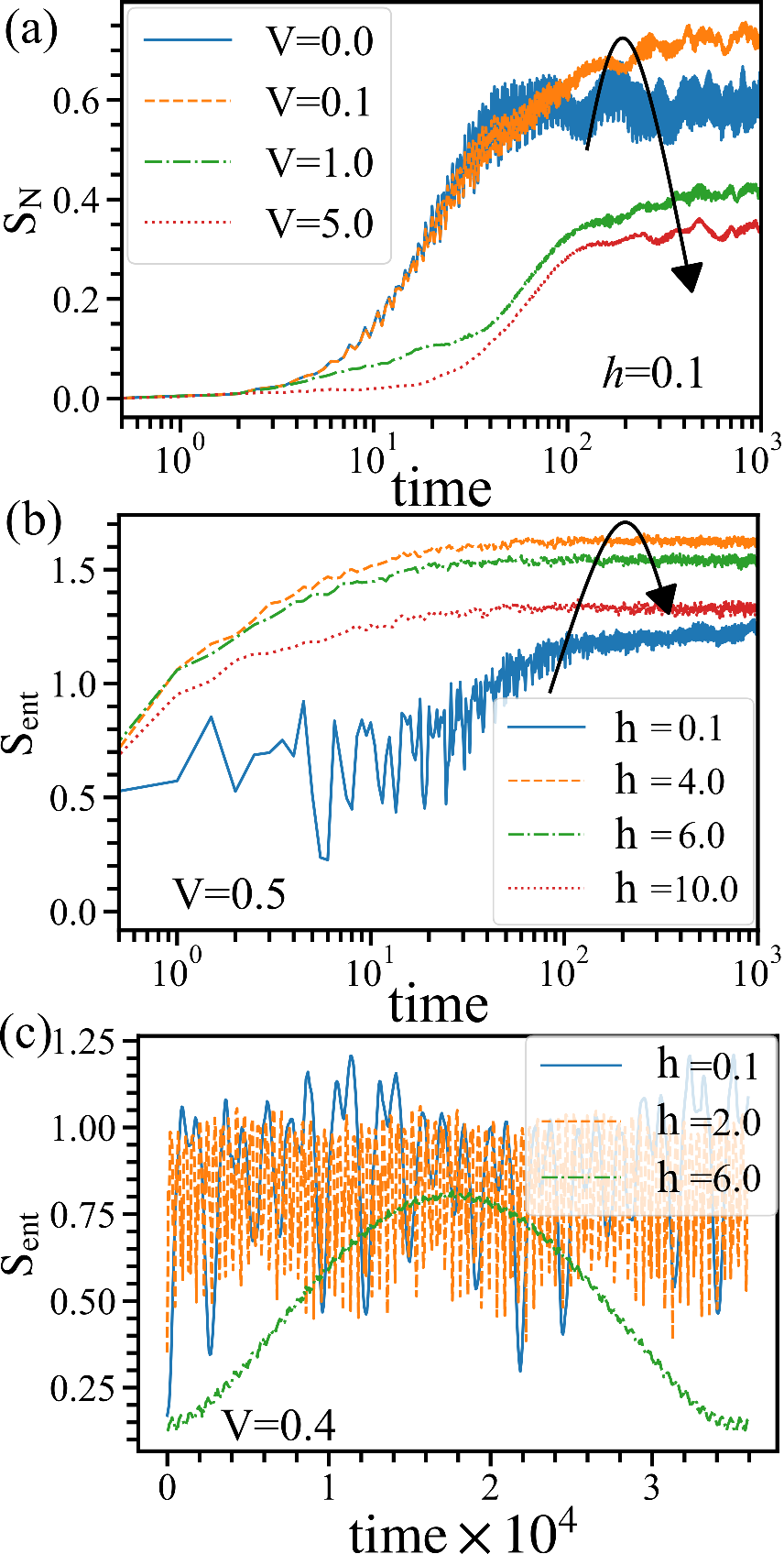} 
\end{center} 
\caption{
(a) Time evolution of $S_{N}$ for $\W=0.1$ with various values of $V$.
(b) Time evolution of $\Se$ for various values of $\W$ for $V=0.5$.
Initial state is $|\psi\rangle_W$.
(c) Time evolution of $\Se$ for $V=0.4$ with various values of $\W$.
Initial state is $|\psi_0\rangle$.
System size $L=10$. We set $\tau_0=\tau_1=1$.
}
\label{Fig2s}
\end{figure}
\begin{figure}[t]
\begin{center} 
\includegraphics[width=7cm]{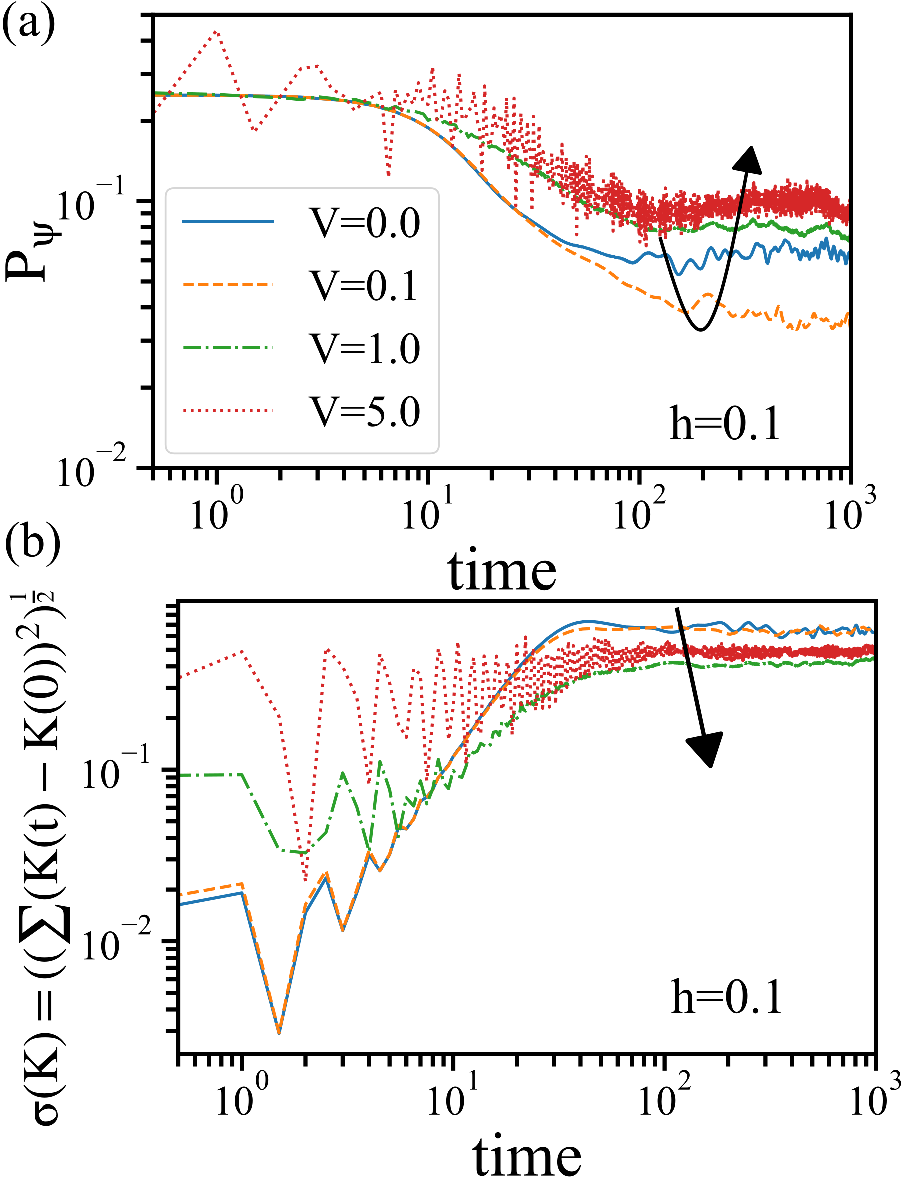} 
\end{center} 
\caption{
Time evolution of $P_\psi$ [(a)] and deviation of $\langle K(t) \rangle$ [(b)] from the initial value for $\W=0.1$
 with various values of $V$.
The data indicate that LIOMs first get unstable in thermal regime and then turn to be stable
in localized regime as $V$ increases further.
}
\label{Fig3s}
\end{figure}

In Figs.~\ref{Fig1s}, we display calculations of the final values of $S_{\rm ent}$, $S_N$
and $P_\psi$ in the time evolution in the $(\W-V)$-plane.
Typical behavior of the time evolution of the physical quantities are shown below.
We find interesting phenomenon for the case of the weak random potentials, i.e., 
weakly interacting system is unstable and exhibits the thermal ergodicity.
As increasing the strength of the interactions, the system tends to enter a MBL state,
which is connected continuously with the FMBL state existing for $\W=0$. 

Figure~\ref{Fig2s} (a) for the time evolution with $\W=0.1$ shows that $S_N$ increases rapidly 
and saturates into a stable value.
The first increase comes from  the AB-caging, and the final value of $S_N$ in the time evolution is a decreasing
function of $V$ except the non-interacting case.
The final values of $V=0.0$ and $0.1$ are close to  $\log 2 \sim 0.69$.
This fact indicates that the state is in the thermal regime.
On the other hand, the small values of $S_{N}$ for the $V=1.0$ and $5.0$ imply
that the interaction enhances localization, which can be regarded as 
the two-particle FMBL.

In Fig.~\ref{Fig2s} (b), we display $\Se$ for $V=0.5$ and various values of $\W$.
We find that $\Se$ increases first as $\W$ increases from $0.1$ to $4.0$, and then $\Se$ decreases 
as $\W$ increases further from $4.0$ to $10.0$.
This behavior indicates that the system enters into the thermal regime first and then reenters to the localized
regime. 
Similar behavior is observed for $S_{N}$.

The above calculations indicate the existence of another type
of localization in the strong random-potential regime besides that coming from 
the CLS by the AB caging in the pure Creutz ladder.
It is expected that localization in that regime is a kind of ordinary MBL, 
which is generated by the interactions between particles localized by Anderson localization.
To verify this expectation, we follow the prescription in Ref.~\cite{Serbyn2012_2} and
calculate $\Se$ by choosing another initial state, which is close to Anderson-localized state. 
That is,
the initial state is a two-particle state as in the above calculation, 
whereas each single-particle
state is a superposition of two localized eigenstates for $V=0$ located closely with each other. 
Explicitly, let us take an initial state given as 
$|\psi_0\rangle={1 \over 2}(|1\rangle+|2\rangle)(|3\rangle+|4\rangle)$,
where $\{|1\rangle, \cdots, |4\rangle\}$ are Anderson-localized states, which are {\it numerically obtained from the single-particle energy eigenstates 
of the non-interacting system} with some specific disorder realization, 
and $\{|1\rangle, |2\rangle\}$($\{|3\rangle, |4\rangle\}$) are neighboring states.
If the interactions are weak and can be treated as a perturbation, the time evolution of
$|\psi_0\rangle$ is determined by the single-particle energies, $\epsilon_\alpha \ (\alpha=1, \cdots,4)$
and the interaction energies, $\delta E_{\alpha\beta}=\langle \alpha| H_{\rm I} |\beta\rangle
\ (\alpha,\beta=1, \cdots,4)$ by assuming the wave functions are not deformed by the interactions.
Then, the straightforward calculation shows that the EE, $\Se$, oscillates with the frequency
$\omega=\delta E_{14}-\delta E_{24}-\delta E_{13}+\delta E_{23}$. As discussed in Ref.~\cite{Serbyn2012_2}, this result comes from the assumption that
the states $\{|1\rangle, \cdots, |4\rangle\}$ are not deformed during the time evolution,
keeping the localized properties for $V=0$. 

Figure.~\ref{Fig2s} (c) shows that $\Se$ exhibits an expected oscillating behavior 
of MBL only for $\W=6.0$, 
where the oscillation frequency is close to $\omega$. This indicates that the Anderson-localized states $\{|1\rangle, \cdots, |4\rangle\}$ 
are not deformed by the interactions during the time evolution.
As discussed in Ref.~\cite{Serbyn2012_2}, this two-particle behavior is nothing 
but a signal
of MBL at finite particle density.
In other words, the increase of $\Se$ for $\W=6.0$ in 
Fig.~\ref{Fig2s} (b) originates
from the fact that the initial state $|\psi\rangle_W$ is far from an energy
eigenstate. 
For $\W=2.0$, on the other hand, $\Se$ oscillates quite rapidly with a fairly large mean value.
This behavior comes from the fact that the state for $\W=2.0$ is an ergodic state 
and  the wave function extends and gets deformed easily because of the large 
overlap between
neighboring states.
Finally for $\W=0.1$, $\Se$ seems to oscillate with multiple frequencies.
In the previous paper \cite{OKI2021}, we studied the system without the disorder and found that $\Se$ exhibits 
a regular oscillation.
We clarified that this oscillation does not come from the mechanism of the energy difference
by the interactions but the mixing of the flat-band localized states in the two-particle sector such as
$|W^+\rangle|W^-\rangle\longleftrightarrow |W^-\rangle|W^+\rangle$.
For $\W=0.1$, the above two oscillation mechanism take place simultaneously, and the multiple
frequencies emerge. 
As a result, a genuine product state does not emerge during the time evolution. 
Anyway, the above observation for Fig.~\ref{Fig2s} (c) clearly indicates the existence of MBL at strong disorders
as well as an ergodic regime for moderate disorder.

\begin{figure*}[t]
\begin{center} 
\includegraphics[width=16cm]{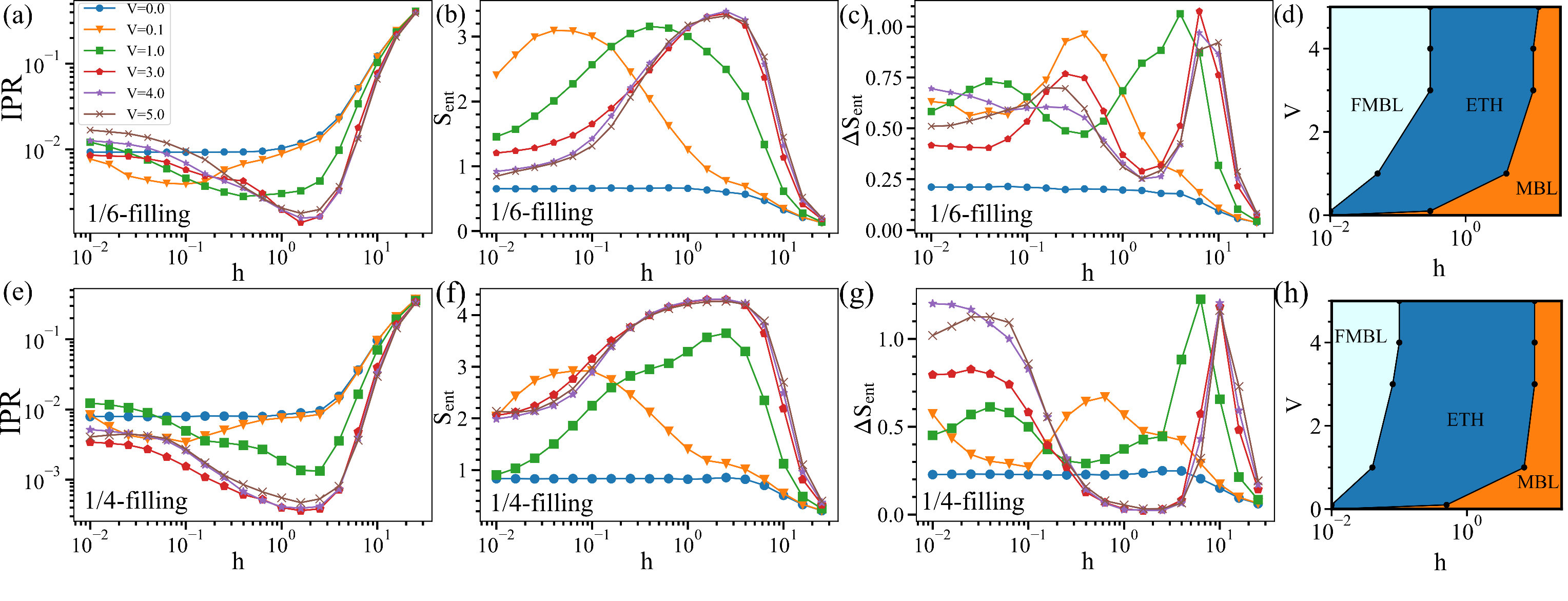} 
\end{center} 
\caption{
IPR, EE and VEE as a function of $\W$ for various $V$'s.
(a)-(c) for the $1/6$-filling case, and (e)-(g) for the $1/4$ filling.
All data indicate the existence of a thermalized (erdogic) state in the intermediate $\W$'s. 
For all data, we picked up the band-center eigenstates. Schematic images of phase structure for the $1/6$-filling [(d)] and $1/4$-filling [(h)], where we use $L=12$ and $L=10$.}
\label{Fig4s}
\end{figure*}

After observing the above behavior of $S_N$, etc., we calculate
physical quantities concerning the LIOMs, i.e., $P_\psi$ and deviation of $\langle K(t)\rangle$ from the initial value
for the weak-disorder system.
The results in Figs.~\ref{Fig3s} indicate that the LIOMs are getting stable as increasing the strength of the interaction $V$.
Figure~\ref{Fig3s} (a) shows the time evolution of $P_\psi$ from the initial state $|\psi\rangle_W$
with $j=4$ and $k=6$.
For the above initial state, $P_\psi(0)=0.25$, and it decreases considerably for $t\gtrsim10$ for all $V$'s. 
It is obvious that $P_\psi(t)$ decreases fastest for $V=0.1$ and the rate of 
the decrease is suppressed as $V$ increases.
The deviation of $\langle K(t)\rangle$ [$\sigma(K)$] in Fig.~\ref{Fig3s} (b) exhibits similar behavior.
For $V=0$ and $V=0.1$, $\sigma(K)$ is almost constant for $t\gtrsim10^2$ and has almost the same value.
As $V$ increases, $\sigma(K)$ is getting smaller indicating the stability of the LIOMs.
The observed behavior of the LIOMs is consistent with that of $\Se$ and $S_{N}$ shown in the above.

In Figs.~\ref{Fig1s}, we give a schematic ``phase diagram'' of two-particle system, which is useful
for study of finite-density phases of the Creutz ladder in the subsequent section.


\section{Systems at finite-filling fractions}

In this section, we shall study the system at finite filling fractions. 
We mostly consider the $1/6$ and $1/4$ fillings.
In order to find phase transitions between the MBLs and thermal states, we first
investigate
static EE properties and inverse participation ratio (IPR), and then, 
we shall study time evolution of the system
at typical parameter regimes for each observed phase. 
We employ the exact diagonalization within the system size that we can numerically
handle (up to $L\leq 12$) \cite{Quspin}.


\subsection{EE, standard deviation of EE and IPR}

We start to numerically study the static EE and IPR in order to see if
the system  at low fillings such as $1/6$ and $1/4$ has a similar
phase diagram with the two-particle system.
To this end, we define a variance of EE (VEE) in addition to the EE and IPR.
The definition of the VEE is given as follows in terms of the EE of $s$-th state
in energy spectrum for disorder realization $r$, 
\be
&&\langle E(s)\rangle \equiv {1 \over N_r}\sum^{\rm disorder}_r E^r_s, 
\ee
where $N_r$ is the number of disorder realizations. 
In the calculation of EE, the subsystem including the unit cells with $j=1,\cdots, L/2$ is used. See Fig.~\ref{model} (b). 
The EE of the finite-density system, $\Se$, and its variance, $\Delta\Se$, are defined
for a state ensemble $\{s\}$ as follows:
\be
&& \Se ={1\over N_s}\sum_s \langle E(s)\rangle, \nonumber \\
&& \Delta\Se ={1 \over {N_sN_r}}\sum_{s,r} (E^r_s-S_{\rm ent})^2,
\label{Sent2}
\ee
where $N_s$ is the number of eigenstates calculated for each disorder realization.
The IPR, $\IP$, is defined as follows:
\be
\IP \equiv {1 \over {N_sN_r}}\sum_{s,r} S^r_s,
\ee
where $S^r_s$ is IPR for state $s$ in disorder realization $r$ defined as 
\be
S^r_s= \sum_j\biggl[|\langle a_j\rangle|^4+|\langle b_j\rangle|^4\biggr],
\label{IPR}
\ee
with the normalization $ \sum_j\biggl[|\langle a_j\rangle|^2+|\langle b_j\rangle|^2\biggr]=1$. 
In the later calculations, we use $N_r=10^2-10^3$ disorder realizations and $N_s=5-50$. 
Also, as for the energy of states $E$, the system energy band is normalized for each disorder realization as
$\epsilon \equiv (E-E_{\rm min})/(E_{\rm max}-E_{\rm min})$ with the minimum (maximum) energy $E_{\rm min}$
($E_{\rm max}$) for the disorder realization. 
We found that the above quantities as a function of $\W$ for fixed values of $V$ exhibit quite
similar behavior for $\epsilon=0.25, 0.5$ and $0.75$.
Hence, in the following we show the results for the band-center states with $\epsilon \sim 0.5$.
In the practical calculation, $N_s\times N_r= 5\times 10^3$.

\begin{figure*}[t]
\begin{center} 
\includegraphics[width=16cm]{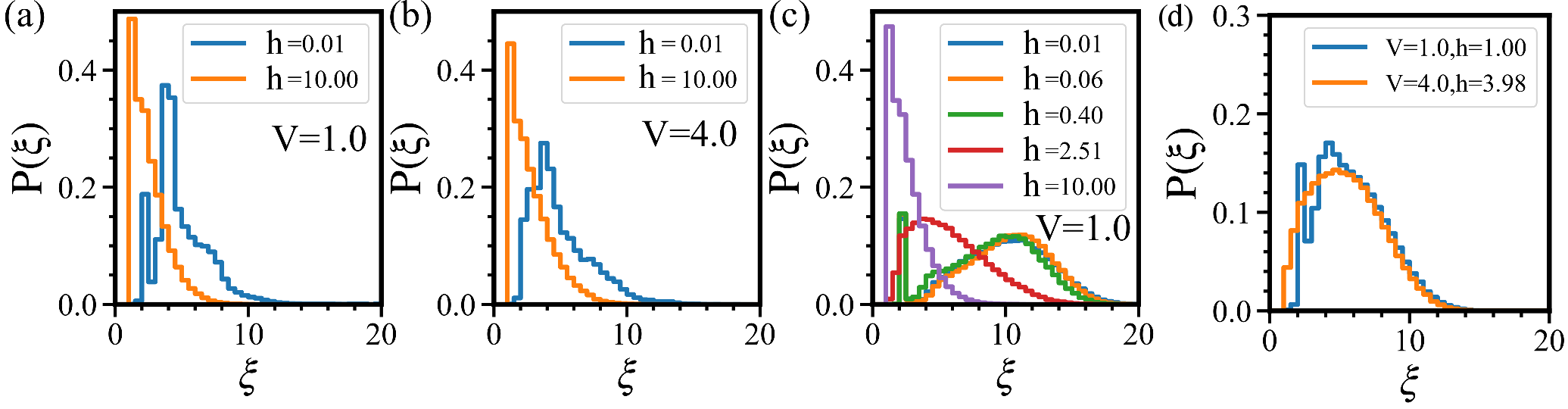} 
\end{center} 
\caption{
Distribution of the LL for various $V$ and $\W$. 
(a) $V=1.0$, $\W=0.01$, $10.0$. For the weak disorder, states are FMBL, and peak of LL 
is $\xi\sim 4$, whereas for the strong disorder, $\xi\sim1$.
For $\W=0.01$, a peak at $\xi\sim 1$ comes from the gapless edge modes representing
symmetry-protected topology.
(b)  $V=4.0$, $\W=0.01$, $10.0$. Edge modes disappear as a result of the strong interaction.
(c) Distribution of LL in a non-flat band system ($v=0.1$).  
LL tends to get shorter as $W$ increases changing from extended to localized states.
(d) Distribution of LL in a thermal regime in the flat band system for moderate $\W$. 
The distribution is broad as in the extended state in (c)
indicating that the system is entirely in extended state.
}
\label{Fig5s}
\end{figure*}

$I_{\rm IPR}$, $\Se$ and $\Delta \Se$ in Figs.~\ref{Fig4s} (a)--(c) show the obtained data for the $1/6$-filling system.
Without random potential ($\W=0$), the calculations of $\Se$ and $\IP$ show 
that the system tends to localize as $V$ increases.
This behavior was observed in our previous work \cite{KOI2020}.
As $\W$ is turned on, the system tends to extend for all $V$'s indicating the presence of thermal regime
for the intermediate values of $\W$.
As $\W$ increases further, both $\Se$ and $\IP$ indicate that the system reenters the localized regime. 
In the whole parameter range of $\W$, the sequential transitions such as ``localized $\to$ thermalized $\to$ localized''
are observed.
This seems somewhat counter intuitive, but has been observed and recognized for other models of MBL.

VEE in Eq.~(\ref{Sent2}) [$\Delta \Se$] shown in Fig.~\ref{Fig4s} (c) is quite useful to characterize 
the critical regimes and phase boundaries. 
For $V=0$, $\Delta \Se$ is a smooth function of $\W$, which indicates that FMBL changes smoothly to MBL 
through a crossover \cite{KOI2020}.
$\Delta \Se$ for $V=0.1$ exhibits a broad peak implying a broad critical regime between FMBL and MBL.
For $V\geq 1.0$,
the obtained $\Delta \Se$ has a double-peak shape, and the location of the second peak slightly
shifts to larger $\W$ as $V$ increases.
The first peak corresponds to the FMBL-thermal 
transition and the second one to thermal-BML by the random potentials. 
[In Appendix A, we show the system-size 
dependence of $\Se$ to examine the thermodynamic limit.]
Therefore, we conclude that
the phase diagram of the system at $1/6$-filling is quite similar to that of the two-particle
system displayed in Fig.~\ref{Fig1s}, and an erdogic phase emerges clearly for $V\geq 0.1$.
This implies that the density of particle in the $1/6$-filling system is low enough, and the
behavior of the two-particle system persists there. Summerizing the data of Fig.~\ref{Fig4s} (a)-(c), we sketch an qualitative phase diagram in Fig.~\ref{Fig4s} (d).
The structure is close to the two-particle case in Fig.~\ref{Fig1s} (a) and (b).

In Figs.~\ref{Fig4s} (e)--(g), we show the calculations of $\Se$, etc. for the $1/4$-filling case.
We find that the $1/4$-filling system behaviors similarly to the $1/6$-filling one.
However in the vicinity of $\W=0$ and $V > 1.0$, the system has rather large $\Se$, and also
$\Delta \Se$ decreases quite rapidly there as $\W$ increases.
Therefore, in the $1/4$-filling case ergodic tendency is more enhanced for $\W\simeq 0$ 
and $V >1.0$ than in the 1/6-filling.
This result indicates that effects of the interactions get stronger as the density 
of particle increases,
and the ergodic nature is enhanced in the $1/4$-filling system compared to the $1/6$-filling one.
This ergodicity enhancement is also observed for the large-$\W$ region. 
Data in Figs.~\ref{Fig4s} (c) and (g) show that the peaks of $\Delta \Se$ are located at larger values
of $\W$ in the $1/4$-filling case compared with those of the $1/6$-filling case. Summerizing the data of Fig.~\ref{Fig4s} (e)-(g), we also sketch an qualitative phase diagram in Fig.~\ref{Fig4s} (h).
The structure is close to that of $1/6$-filling and the two-particle case in Fig.~\ref{Fig1s} (a) and (b).

\subsection{One particle density matrix analysis}
We would like to characterize the two kinds of MBL
at $1/6$-filling in Figs.~\ref{Fig4s} (a)--(c), and then, 
we investigate the localization length (LL) by using the methods of the one-particle-density-matrix (OPDM)
proposed in Ref.\cite{Bera,Bera2017}. 
The OPDM, $\hat{\rho}$, for a many-particle wave function, $|\psi\rangle$, is defined as
\be
\rho_{ij}=\langle \psi|c^\dagger_ic_j|\psi\rangle,
\label{SPDM1}
\ee
where we have renumbered sites of the ladder: odd (even) sites correspond to the upper (lower) leg on which 
$a$-particles ($b$-particles) reside, and $c_i=a_i \ (b_i)$ for odd (even) $i$.
Then, natural orbits, $|\phi_\alpha\rangle$ $(\alpha=1, \cdots, 2L)$, are obtained by solving equations,
\be
\hat{\rho}|\phi_\alpha\rangle=n_\alpha|\phi_\alpha\rangle,
\label{NO}
\ee
where $\{ n_\alpha\}$ are positive eigenvalues and $n_1 <n_2<\cdots<n_{2L}$.
The total particle number $N$ of $|\psi\rangle$ is related with $\{n_\alpha\}$ as 
$\sum^{2L}_{\alpha=1} n_\alpha=N$.

\begin{figure}[t]
\begin{center} 
\includegraphics[width=9cm]{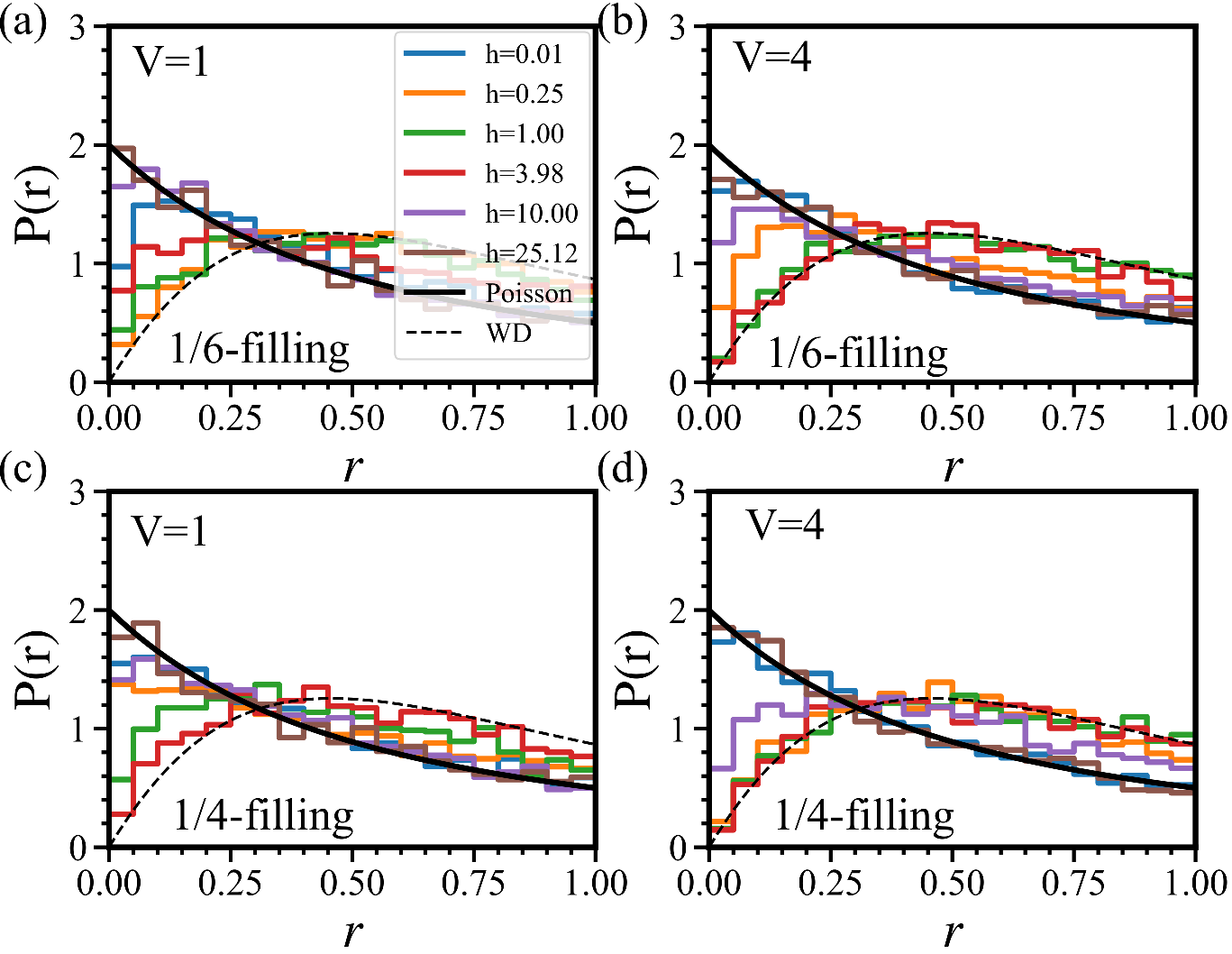} 
\end{center} 
\caption{Level-statistical distributions with fixed interactions ($V=1$ and $V=4$) 
for various disorder strengths. 
(a) and (b): $1/6$-filling cases. (c) and (d): $1/4$-filling cases. 
The black solid (dashed) line represents the Poisson (Wigner-Dyson) distribution.
We used 100 disorder realizations.}
\label{gap_ratio}
\end{figure}

By using the normalized natural orbits, $\{|\phi_\alpha\rangle \}$, we define the following `single-particle IPR':
\be
I^{\alpha}_{\rm SP}=\sum^{2L}_{i=1}|\phi_\alpha(i)|^4.
\label{SPIPR}
\ee
The LL of single particles in the many-body state $|\psi\rangle$, $\xi^{\alpha}$, is naturally identified as 
$\xi^{\alpha}=1/I^{\alpha}_{\rm SP}$. 
In the previous paper \cite{KOI2020}, we calculated the LL from the IPR in Eq.~(\ref{IPR}).
The above definition of LL is more acceptable as it is directly related to the natural orbits.

We consider the $1/6$-filling system.
In Figs.~\ref{Fig5s} (a) and (b), we plot the distribution of the LL of states in the central range of the energy spectrum.
For both $V=1.0$ and $4.0$ cases, the strong-disordered states have short LL, whereas the weak-disordered states
have relatively long LL.
Peaks of the LL distribution for $\W=0.01$ are located at $\xi\sim 4$ for both $V=1.0$ and $4.0$, and this result
implies that the flat-band LIOMs dominate over the interactions and disorder. This observation in terms of the LL clearly characterizes two kinds of localization;
one of the flat-band CLS type and the other of ordinary MBL. 

It is interesting to see how the above obtained results of localization
are influenced by a change in the single-particle band structure.
In particular, flat-band localization realized for small $\W$ is supported by the CLS and therefore,
it may disappear for a perturbation breaking the flat band. 
Figure~\ref{Fig5s} (c) shows the distribution of the LL of a non-flat system, 
in which an inter-leg
hopping term, $H_{\rm ILH}=-v\sum_j(a^\dagger_jb_j+\mbox{h.c.})$ is added to make the band dispersive. 
Except for $\W=10$ case, the LL tends to be large, indicating that the system is in extended regime.
As $\W$ increases, the LL tends to get shorter. We observed similar behavior of the LL for other values of $V$ and $v$.
This result obviously indicates that the flat-band localized regime disappears by the existence of the inter-leg hopping as it makes the single-particle energy spectrum dispersive.

For $\W\leq 0.4$, we observe an isolated peak of $P(\xi)$ at $\xi\sim 2$ in both Figs.~\ref{Fig5s} (a) and (c), 
which indicates the existence of the edge modes.
We expect that this result comes from the fact that the genuine Creutz ladder without interactions and disorder is 
a symmetry-protected topological (SPT) model. With open boundary, the gapless edge mode is obtained by cutting the single CLS residing on the edge, the form of which is given by $(a_1+ib_1)$.
More detailed study on this topological feature of the present system
is a future work. 

Furthermore, we observe the thermal regime in detail. 
For the typical parameter sets of the thermal phase, 
we plot the distribution of LL in Fig.~\ref{Fig5s} (d).
The distribution of LL is broad, and specific peaks do not appear
similarly to the thermal phase of $v=0.1$ and small values of $\W$. 
This indicates that the system is in extended phase, that is thermal ETH phase.
This observation will be confirmed by the investigation of the energy-level statistics
in the following subsection. 

\begin{figure*}[t]
\begin{center} 
\includegraphics[width=14cm]{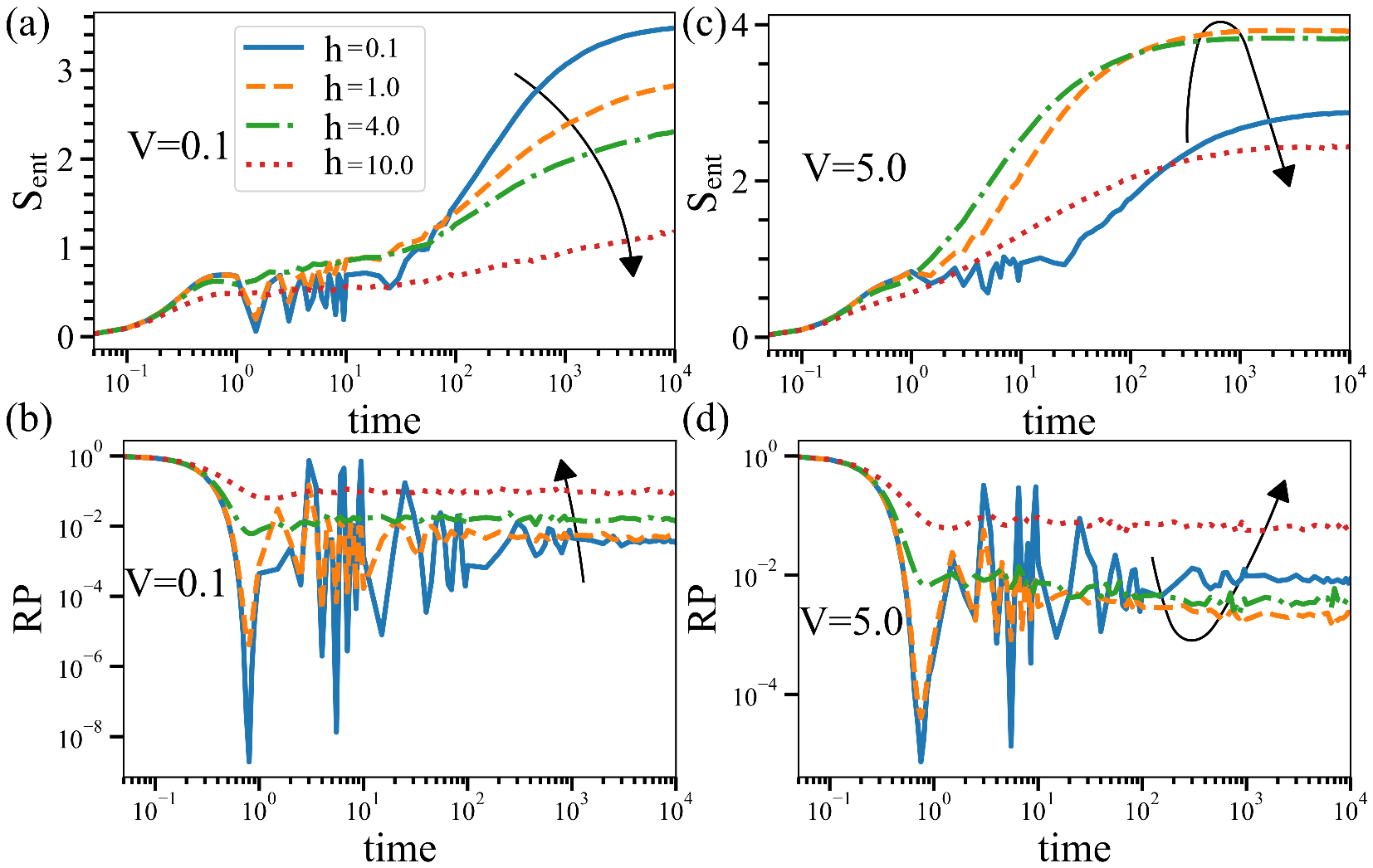} 
\end{center} 
\caption{
Time evolution of the $1/6$-filling state.
(a) EE as a function of $\W$ for $V=0.1$.
(b) Return probability  as a function of $\W$ for $V=0.1$.
(c) EE as a function of $\W$ for $V=5.0$.
(d) Return probability  as a function of $\W$ for $V=5.0$.
The arrows indicate how the physical quantities change as $\W$ increases.
All calculations are in good agreement with the data of IPR, EE and VEE shown in Fig.~\ref{Fig4s}.
In the $V=5.0$ state, the system exhibits ``FMBL $\to$ ergodic $\to$ MBL'' transitions.
}
\label{Fig6s}
\end{figure*}

\subsection{Detailed investigation for ergodic regime by using the level statistics}

From the calculations of EE, IPR, and LL extracted by the OPDM, the three phases, 
that is, FMBL, thermal, and MBL are expected to exist in the present system. 
Here, we further characterize each phases and especially elucidate the presence of 
the thermal regime between the two kinds of MBL phases. 
To this end, we study the level statistics \cite{Oganesyan2007}.

We impose the periodic boundary condition, and obtain all energy eigenvalues of 
the system, $E_k$, which are sorted in ascending order.  
Then, we calculate the level spacing $r_k$ defined by $r_{k}=[{\rm min}(\Delta^{(k)}, \Delta^{(k+1)})]/[{\rm max}(\Delta^{(k)},\Delta^{(k+1)})]$ for all $k$, where $\Delta^{(k)}=E_{k+1}-E_{k}$, and observe the statistical distribution of $r_k$ after averaging over disorder realizations. 
[In the practical calculation, we used 4800 samples.]
If the distribution is close to the Wigner-Dyson (WD) distribution, 
the system is ergodic and thermal, 
while if close to the Poisson distribution, the system is in a localized regime. 

We calculated the level statistics for the case with moderate ($V=1$) and strong
interactions ($V=4$) \cite{Come_small_V}. 
The $1/6$-filling case for various $\W$'s are shown in Fig.~\ref{gap_ratio} (a) and (b). 
We observe that for small and large-$h$ cases, the distributions are close to the Poisson
distribution, on the other hand, for moderate $h$, the distribution is close to 
the WD distribution. 
This result obviously indicates the presence of the thermal (ETH) phase for 
moderate $\W$. 
The same observation is obtained for the $1/4$-filling case as shown in
Fig.~\ref{gap_ratio} (c) and (d). 
These results of Fig.~\ref{gap_ratio} are consistent with the results of the EE, IPR,
and the LL obtained as in Figs.~\ref{Fig4s} and \ref{Fig5s}. 

\subsection{Time evolution of systems at finite fillings}

In this subsection, we shall observe the time evolution of the system at filling fractions $1/6$ and $1/4$.
In the previous subsection, we obtained the schematic overview of the phase diagram, and then,
we shall focus on the system with $V=0.1$ and $5.0$ for various values of $\W$. 
Data in Figs.~\ref{Fig4s} (a)--(c) show that the system for $V=0.1$ at $1/6$ filling fraction
has a rather large ergodic (or critical) regime for 
intermediate values of $\W$, whereas for $V=5.0$, the system changes from FMBL, ergodic and MBL regimes
as $\W$ increases.
We shall see how the system evolves reflecting this phase structure.

\begin{figure}[t]
\begin{center} 
\includegraphics[width=8cm]{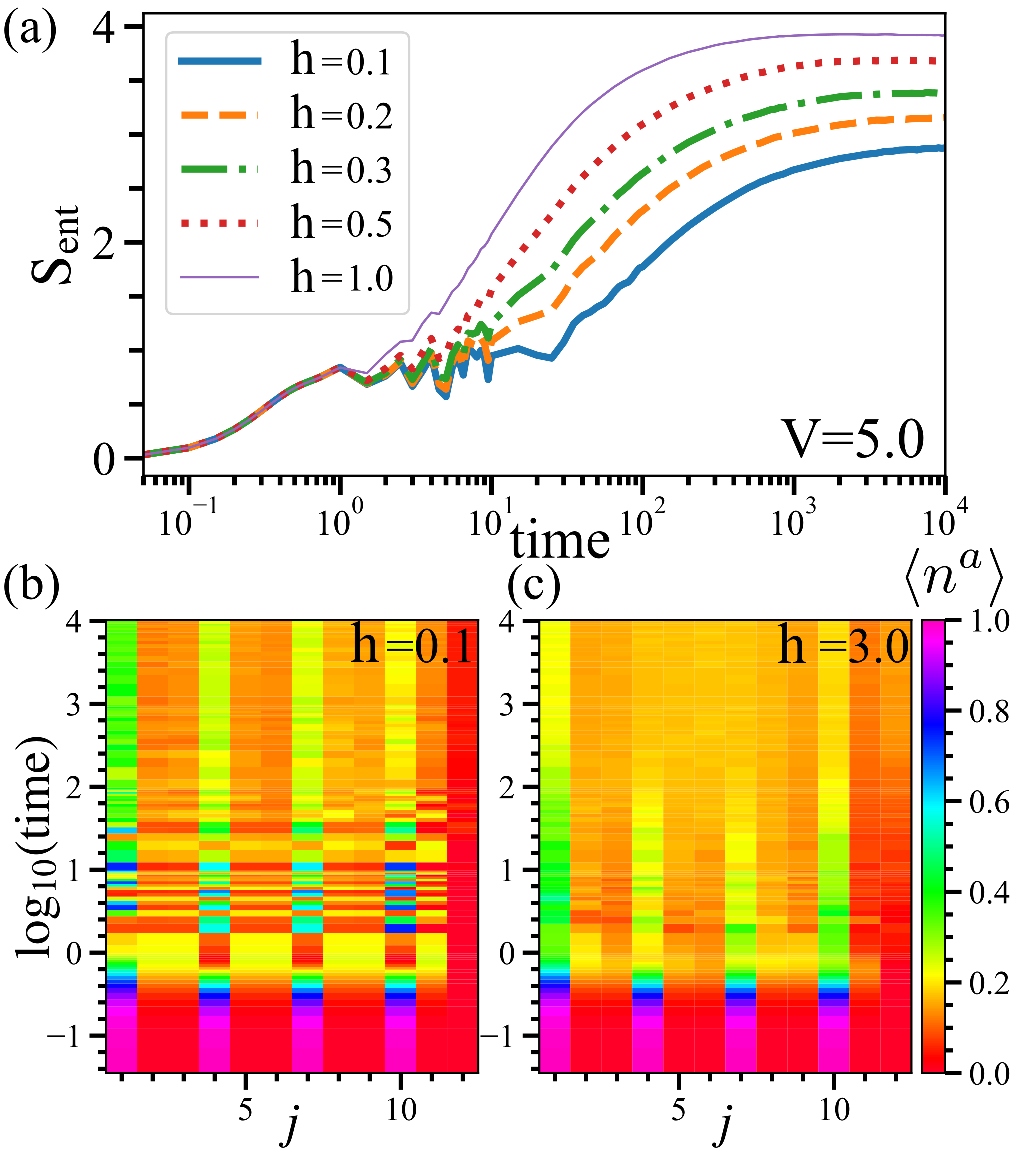} 
\end{center} 
\caption{
(a) Time evolution of EE in the $1/6$-filling state with $V=5.0$ and various $\W$'s.
For smaller value of $\W$, in particular for $\W=0.1$, increase of EE starts at later time.
This behavior comes from the flat-band properties of the states, i.e., the localization length
of the states in FMBL is longer than that of MBL.
Density profile of $a$-particle for $\W=0.1$ [(b)] and $\W=3.0$ [(c)].
}
\label{Fig7s}
\end{figure}
\begin{figure}[t]
\begin{center} 
\includegraphics[width=8cm]{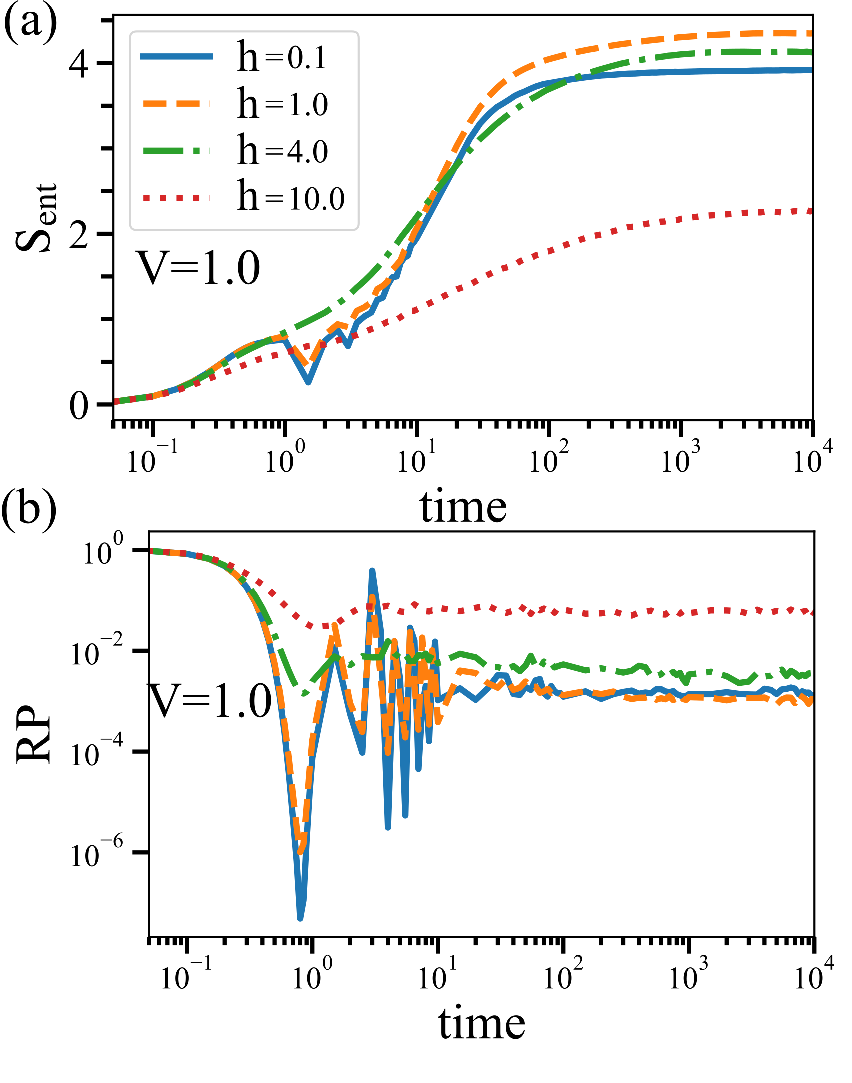} 
\end{center} 
\caption{
Time evolution of the $1/4$-filling state.
(a) EE as a function of $\W$ for $V=1.0$.
(b) Return probability  as a function of $h$ for $V=1.0$.
All calculations are in good agreement with the data of IPR, EE and VEE shown in Fig.~\ref{Fig4s}.
}
\label{Fig8s}
\end{figure}

Firstly, we investigate the $1/6$-filling case, where the initial state is 
prepared as $|\psi(0)\rangle = \prod^{L/3-1}_{n=0}a^\dagger_{3n} |0\rangle$ 
where $L/3$ is an integer. 
Figures~\ref{Fig6s} (a) and (b) display the EE and return probability (RP) for $V=0.1$ system, respectively,
where the RP, $R_{\rm p}(t)$, is defined as $R_{\rm P}(t)= |\langle \psi(t)|\psi(0)\rangle|^2$ with the many-body
wave function at time $t$, $|\psi(t)\rangle$.
The increase (decrease) of the EE (RP) for $10^{-1} \sim 10^0$ comes from the AB caging of the flat-band
nature of each particle, and therefore this behavior is clearer for smaller $\W$.
The time evolution of the $V=0.1$ system for $t>10^1$ shows that the MBL properties emerge for 
the strong disorder case with $\W=10.0$, whereas the other smaller $h$ cases are in the thermalized
regime.
We cannot deny the possibility that these cases have subdiffusive (critical) nature, but our numerical
study of the genuine extended state with a finite inter-leg hopping shows that their time evolutions are close to
that of the extended state.
Anyway, the above time-evolution behavior of the $V=0.1$ case is in good agreement 
with the static localization properties obtained in Fig.~\ref{Fig4s}.

For the $V=5.0$ case, on the other hand, Figs.~\ref{Fig6s} (c) and (d) show that the system with
$\W=0.1$ and $10$ are in the MBL regimes, whereas systems with the intermediate $h$'s have
the thermalized properties.
This result exhibits the existence of the FMBL and MBL for weak and strong disorders, respectively.
Again, the behavior of the time evolution of the $V=5.0$ system is consistent with 
its static properties displayed in Fig.~\ref{Fig4s}.
Furthermore by a close look at the data of $\W=0.1$ and $\W=10$, we find differences in the EE and RP behaviors
for these cases.
That is, for $t=10^1 - 10^2$, the EE of $\W=0.1$ is smaller than that of $\W=10$, and then after $t=10^2$,
the system with $\W=0.1$ starts to extend more rapidly than the system with $\W=10$.  
Obviously, this comes from the difference of the origin of localization in two cases.

In Fig.~\ref{Fig7s} (a), we display the detailed behavior of the EE 
as a function of time for $V=5.0$ and various values of $\W$.
We can see the typical behavior of the AB caging for $t=10^1 - 10^2$ in the system with weak disorders
such as $\W=0.1$ and $0.2$, but not for larger $\W$'s.
This behavior causes a delay in the development of EE.
To verify this,  we display the density profiles in the time evolution for $\W=0.1$ 
and $3.0$ in Figs.~\ref{Fig7s} (b) and (c).
These data clearly show that in the $\W=0.1$ system, the initial density of 
$a$-particle oscillates for a whole
and then it starts to get homogeneous.
On the other hand in the system with strong disorder $\W=3.0$, no oscillations are observed
indicating the absence of the AB caging.

Finally, in Fig.~\ref{Fig8s}, we show the EE and RP for the $1/4$-filling fraction.
These quantities for $V=1.0$ exhibit similar but more erogodic behavior compared with those in the $1/6$-filling case,
i.e., the system with $\W=10$ exhibit the MBL nature but the others with smaller $\W$ are in the ergodic state.
This implies that the interactions work more effectively in higher density systems to enhance thermalization
as in the ordinary disordered systems.
Qualitatively similar properties are observed for the $V=5.0$ case, but the ergodic nature is enhanced more 
due to the interactions. [These numerical results are shown in Appendix B.]
This is again consistent with data in Fig.~\ref{Fig4s}.

\section{Conclusion and discussion}

In this paper, we studied the Creutz ladder with interactions and random potentials.
The pure Creutz ladder has two dispersionless-flat bands where all states are 
localized due to the presence of the CLS, which are finite-support localized states.
The LIOMs are explicitly obtained by the CLS;
the Creutz ladder is an example of the disorder-free localized system.
On the other hand by applying random potentials, 
it is expected that ordinary MBL localization by the strong disorder
emerges.
Therefore, the target model is one of the ideal systems for investigating the interplay and competition
between disorder, interactions and flat band.
Furthermore, phase transitions between two distinct MBL states
have been less studied compared with MBL-thermal transitions, and a few works 
on it appeared recently \cite{Sahay,MHK}.

In this work, we first clarified the phase diagram of the two-particle system by investigating the time evolution
of the EE and LIOMs. 
We found that there exist two kinds of MBL, one corresponds to FMBL by the AB caging
and the other to MBL by random potentials.
Between these localized regimes, an erdogic state emerges, which is enhanced by 
the interactions.

Next, we studied localization properties of the system at finite filling fractions, i.e., $1/6$ and $1/4$ fillings.
By using the obtained two-particle phase diagram, we focused on parameter ranges, which are expected to be
physically significant.
Static localization properties were investigated by calculating EE, VEE, and IPR
for $V=0.1$ and $5.0$ with various $\W$'s. 
In particular, we found a double peak structure in $\Delta \Se$ as increasing disorder. 
This indicates the existence of two kinds
of MBL and thermalized regime between them.
Furthermore, study of the time evolution of the EE and RP and the above static properties clarify 
the localization properties of the system for each fraction, and reveals the existence of two kinds
of MBL and thermalized regime between them. 
This behavior was also characterized by the level statistics.
The $1/4$-filling state is more ergodic that the $1/6$-filling one, as the interactions work more efficiently in a higher-density state.

Comments are in order.
In addition to the repulsive interactions in $H_{\rm I}$ in Eq.~(4), 
a repulsion such as $\sum_j(n^a_jn^b_{j+1}+n^a_{j+1}n^b_j)$ may appear in real experiments. 
This additional repulsion generates similar effects on localization as $H_{\rm I}$. 
This observation has been supported in our previous work \cite{KOI2020}.

In the previous work on the $1/8$-filling system \cite{KOI2020}, we found that there emerge many-body states in which FMBL and 
ordinary MBL take place simultaneously for fixed values of $V$, $\W$ and filling fraction ($1/8$).
More precisely, states in the vicinity of edges of the energy band exhibit the FMBL nature, whereas states in the 
central regime of the energy spectrum are localized by the disorder potentials.
In Fig.~\ref{Fig5s}, we display the distribution of the LL solely in the band-center regime.
It is interesting to investigate that a similar `hybrid localization' takes place in the system at various
fillings.
Furthermore, there are a few flat-band systems besides the Creutz ladder, and some of them
were studied recently.
It is a future problem to study if a hybrid localization is a generic phenomenon or specific one.

\section*{ACKNOWLEDGMENTS}
The work is supported by JSPS
KAKEN-HI Grant Number JP21K13849 (Y.K.).
\appendix

\section{System size dependence of EE and IPR}
\begin{figure}[t]
\begin{center} 
\includegraphics[width=6cm]{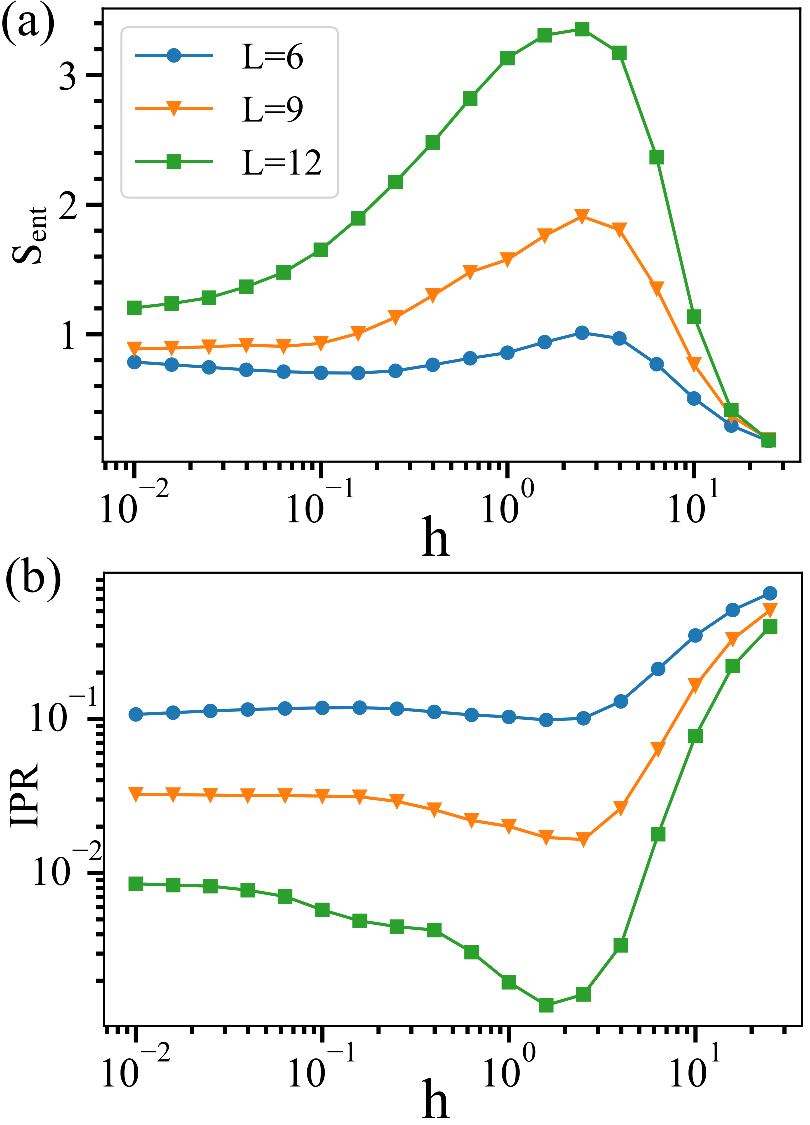} 
\end{center} 
\caption{System-size dependence of of the $1/6$-filling state.
(a) EE as a function of $\W$ for various system size $L$.
(b) IPR as a function of $\W$ for various system size $L$.
}
\label{FigAA}
\end{figure}

In this appendix, we show numerical study of the system-size dependence of 
EE and IPR.
Figures~\ref{FigAA} clearly indicate that both EE and IPR have expected system size 
dependence.
Therefore, we expect that the ergodic regime exists between the FMBL and MBL states
in the thermodynamic limit. 

\begin{figure}[h]
\begin{center} 
\includegraphics[width=6cm]{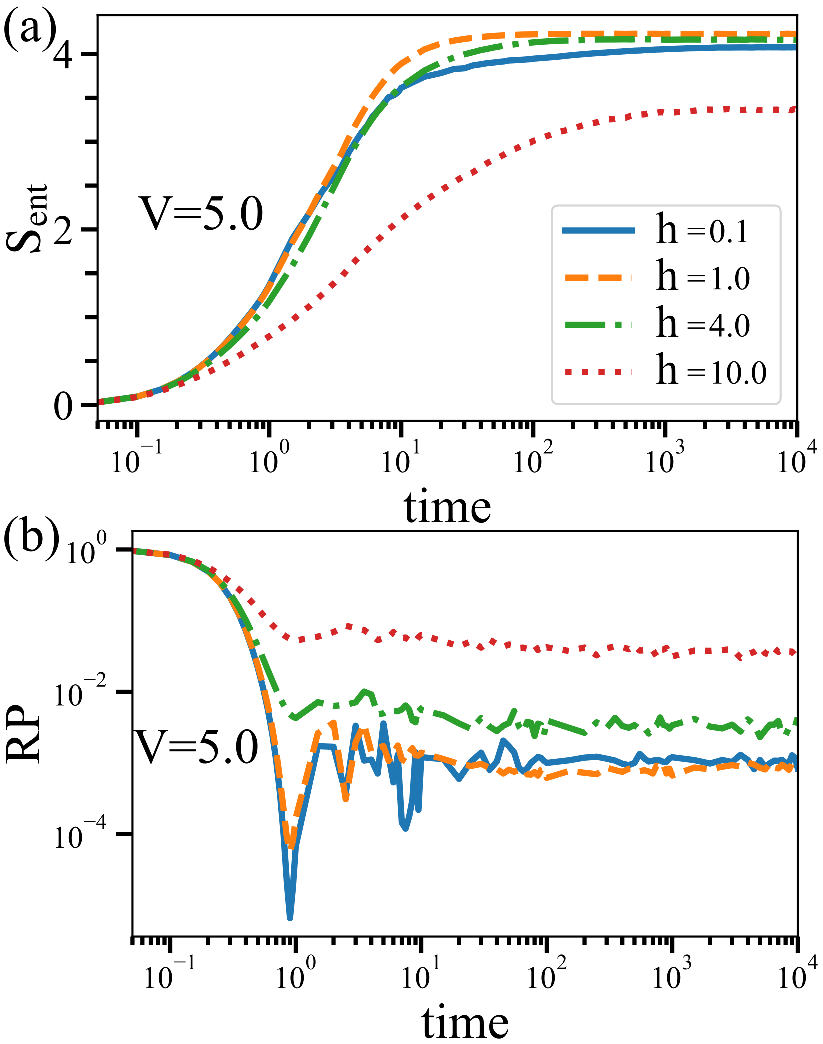} 
\end{center} 
\caption{
Time evolution of the $1/4$-filling state.
(a) Time evolution of EE for various $\W$'s.
(b) Time evolution of RP for various $\W$'s.
}
\label{FigAB}
\end{figure}
\section{Time evolution of $1/4$-filling system with $V=5.0$}

In Fig.~\ref{Fig8s} in Sec.~IV B, we showed the time evolution of the $1/4$-filling
state with $V=1.0$.
In this appendix, we show the results of the $V=5.0$ system in Fig.~\ref{FigAB}.
Qualitatively, the above two systems are close with each other, but there
exist certain differences in numerical results.
That is, the ergodic regime in the $V=5.0$ system is larger than that in the $V=1.0$ system.
The inter-particle repulsions enhance delocalization as in the usual cases.

\newpage


\begin{thebibliography}{99}
\bibitem{Come_small_V}
For very small $V$ case, it is difficult to extract reliable data of the level statistics
due to the effects of degeneracy. Other methods are preferred, which may be a future work.
\end{thebibliography}
\end{document}